\documentclass[10pt]{article}

\usepackage{amsmath}
\usepackage{amssymb}

\usepackage{graphicx}

\usepackage[comma,sort&compress,square]{natbib}

\usepackage{color}

\usepackage[labelfont=bf,labelsep=period,justification=raggedright]{caption}

\makeatletter
\renewcommand{\@biblabel}[1]{\quad#1.}
\makeatother

\date{}

\DeclareMathOperator{\rms}{rms}
\newcommand{\cell}{\text{cell}}
\newcommand{\bigO}{\mathcal{O}}

\usepackage[breaklinks]{hyperref}
\usepackage[version=3]{mhchem}
\usepackage{fixltx2e,hypernat}

\begin{document}

\begin{flushleft}
{\Large
\textbf{Bistability in Apoptosis by Receptor Clustering}
}
\\
Kenneth L. Ho$^{1, \ast}$,
Heather A. Harrington$^{2}$
\\
\bf{1} Courant Institute of Mathematical Sciences and Program in Computational Biology, New York University, New York, New York, United States of America
\\
\bf{2} Department of Mathematics and Centre for Integrative Systems Biology at Imperial College, Imperial College London, London, United Kingdom
\\
$\ast$ E-mail: \url{ho@courant.nyu.edu}
\end{flushleft}

\section*{Abstract}
Apoptosis is a highly regulated cell death mechanism involved in many physiological processes. A key component of extrinsically activated apoptosis is the death receptor Fas, which, on binding to its cognate ligand FasL, oligomerize to form the death-inducing signaling complex. Motivated by recent experimental data, we propose a mathematical model of death ligand-receptor dynamics where FasL acts as a clustering agent for Fas, which form locally stable signaling platforms through proximity-induced receptor interactions. Significantly, the model exhibits hysteresis, providing an upstream mechanism for bistability and robustness. At low receptor concentrations, the bistability is contingent on the trimerism of FasL. Moreover, irreversible bistability, representing a committed cell death decision, emerges at high concentrations, which may be achieved through receptor pre-association or localization onto membrane lipid rafts. Thus, our model provides a novel theory for these observed biological phenomena within the unified context of bistability. Importantly, as Fas interactions initiate the extrinsic apoptotic pathway, our model also suggests a mechanism by which cells may function as bistable life/death switches independently of any such dynamics in their downstream components. Our results highlight the role of death receptors in deciding cell fate and add to the signal processing capabilities attributed to receptor clustering.

\section*{Author Summary}
Many prominent diseases, most notably cancer, arise from an imbalance between the rates of cell growth and death in the body. This is often due to mutations that disrupt a cell death program called apoptosis. Here, we focus on the extrinsic pathway of apoptotic activation, which is initiated upon detection of an external death signal, encoded by a death ligand, by its corresponding death receptor. Through the tools of mathematical analysis, we find that a novel model of death ligand-receptor interactions based on recent experimental data possesses the capacity for bistability. Consequently, the model supports threshold-like switching between unambiguous life and death states---intuitively, the defining characteristic of an effective cell death mechanism. We thus highlight the role of death receptors, the first component along the apoptotic pathway, in deciding cell fate. Furthermore, the model suggests an explanation for various biologically observed phenomena, including the trimeric character of the death ligand and the tendency for death receptors to colocalize, in terms of bistability. Our work hence informs the molecular basis of the apoptotic point-of-no-return, and may influence future drug therapies against cancer and other diseases.

\section*{Introduction}
Apoptosis is a coordinated cell death program employed by multicellular organisms that plays a central role in many physiological processes. Normal function of apoptosis is critical for development, tissue homeostasis, cell termination, and immune response, and its disruption is associated with pathological conditions such as developmental defects, neurodegenerative disorders, autoimmune disorders, and tumorigenesis \cite{thompson:1995:science,raff:1998:nature,meier:2000:nature,fulda:2006:oncogene,taylor:2008:nat-rev-mol-cell-biol}. Due to its biological significance, much effort has been devoted to uncovering the pathways governing apoptosis. Indeed, recent progress has enabled the proliferation of mathematical models, both mechanistic and integrative \cite[e.g.,][]{fussenegger:2000:nat-biotechnol,bentele:2004:j-cell-biol,eissing:2004:j-biol-chem,hua:2005:j-immunol,bagci:2006:biophys-j,legewie:2006:plos-comput-biol,albeck:2008:mol-cell,albeck:2008:plos-biol,okazaki:2008:j-theor-biol}, which together have offered profound insights into the underlying molecular interactions. The current work takes a similarly mathematical approach and hence inherits from this legacy.

There are two main pathways of apoptotic activation: the extrinsic (receptor-mediated) pathway and the intrinsic (mitochondrial) pathway, both of which are highly regulated \cite{budihardjo:1999:annu-rev-cell-dev-biol,danial:2004:cell}. In this study, we focus on the core machinery of the extrinsic pathway, which is initiated upon detection of an extracellular death signal, e.g., FasL, a homotrimeric ligand that binds to its cognate transmembrane death receptor, Fas (CD95/Apo-1), in a 1:3 ratio. This clusters the intracellular receptor death domains and promotes the ligation of FADD, forming the death-inducing signaling complex (DISC) \cite{ashkenazi:1998:science,peter:1998:curr-opin-immunol,peter:2003:cell-death-differ}. The DISC catalyzes the activation of initiator caspases, e.g., caspase-8, through death effector domain interactions. Initiator caspases then activate effector caspases, e.g., caspase-3, which ultimately execute cell death by direct cleavage of cellular targets \cite{nicholson:1997:trends-biochem-sci,nunez:1998:oncogene,thornberry:1998:science,nicholson:1999:cell-death-differ}.

Apoptosis is typically viewed as a bistable system, with a sharp all-or-none switch between attracting life and death states. This bistability is important for conferring robustness \cite{kitano:2004:nat-rev-genet}. Consequently, researchers have used computational models to identify and study potential sources of bistability in apoptosis, including positive caspase feedback \cite{eissing:2004:j-biol-chem}, inhibition of DISC by cFLIP \cite{bentele:2004:j-cell-biol}, cooperativity in apoptosome formation \cite{bagci:2006:biophys-j}, double-negative caspase feedback through XIAP \cite{legewie:2006:plos-comput-biol}, and double-negative feedback in Bcl-2 protein interactions \cite{cui:2008:plos-one}. In this work, we propose that bistability may be induced upstream by the death receptors themselves.

The current model of death ligand-receptor dynamics assumes that FasL activates Fas by direct crosslinking, producing a DISC concentration that varies smoothly with the ligand input \cite{lai:2004:math-biosci-eng}. However, recent structural data \cite{scott:2009:nature} suggests a different view. In particular, Fas was found in both closed and open forms, only the latter of which allowed FADD binding and hence transduction of the apoptotic signal. Moreover, open Fas were observed to pair-stabilize through stem helix interactions. This affords a mechanism for bistability, similar to the Ising model in ferromagnetism \cite{ising:1925:z-phys}, where open Fas, presumably disfavored relative to their native closed forms \cite{huang:1996:nature}, are able to sustain their conformations even after removal of the initial stimulus promoting receptor opening, past a certain critical density of open Fas. This induces hysteresis in the concentration of active, signaling receptors and therefore in apoptosis.

\begin{figure}[!ht]
 \includegraphics{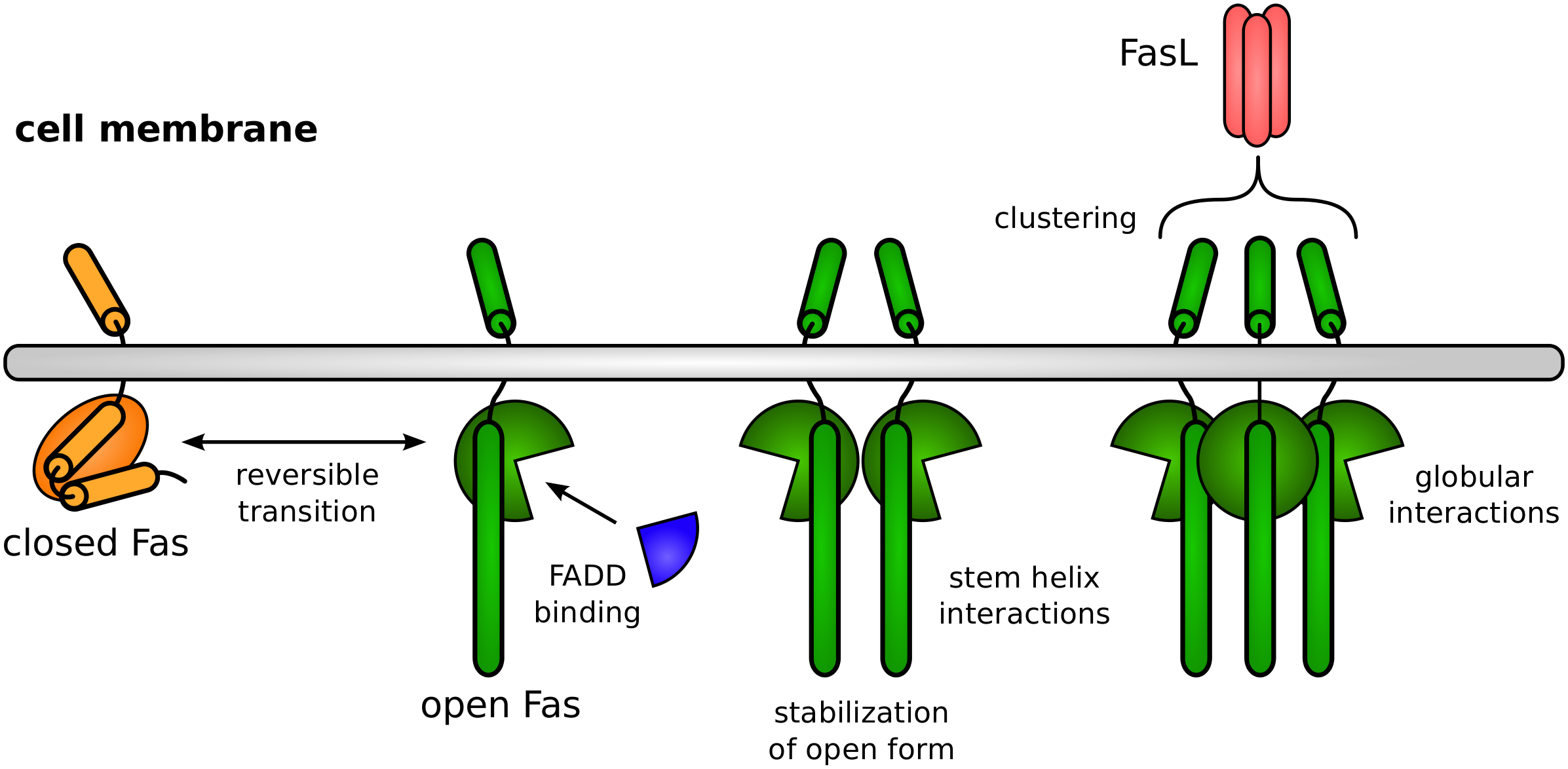}
 \caption{{\bf Cartoon of model interactions.} The transmembrane death receptor Fas natively adopts a closed conformation, but can open to allow the binding of FADD, an adaptor molecule that facilitates apoptotic signal transduction. Open Fas can self-stabilize via stem helix and globular interactions, which is enhanced by receptor clustering through association with the ligand FasL.}
 \label{fig:1}
\end{figure}

We studied this proposed mechanism by formulating and analyzing a mathematical model. The essential interpretation is that FasL acts as a clustering platform for Fas, which establish contacts with other Fas through pairwise and higher-order interactions to form units capable of hysteresis (Figure \ref{fig:1}). At low receptor concentrations, the model exhibits bistability provided that the number of receptors that each ligand can coordinate is at least three. This hence gives a theory for the trimeric character of FasL. Furthermore, at high concentrations, for example, through receptor pre-association \cite{chan:2000:science,siegel:2000:science,chan:2007:cytokine} or localization onto lipid rafts \cite{muppidi:2004:nat-immunol}, irreversible bistability is achieved, implementing a permanent cell death decision. Thus, our model suggests a primary role for death receptors in deciding cell fate. Moreover, our results offer novel functional interpretations of ligand trimerism and receptor pre-association and localization within the unified context of bistability.

\section*{Results}

\subsection*{Model formulation}
Constructing a mathematical model of Fas dynamics is not entirely straightforward as receptors can form highly oligomeric clusters \cite{muppidi:2004:nat-immunol,scott:2009:nature}. A standard dynamical systems description would therefore require an exponentially large number of state variables to account for all combinatorial configurations. To circumvent this, we considered the problem at the level of individual clusters. Each cluster can be represented by a tuple denoting the numbers of its molecular constituents, the cluster association being implicit, so only these molecule numbers need be tracked.

\begin{figure}[!ht]
 \includegraphics{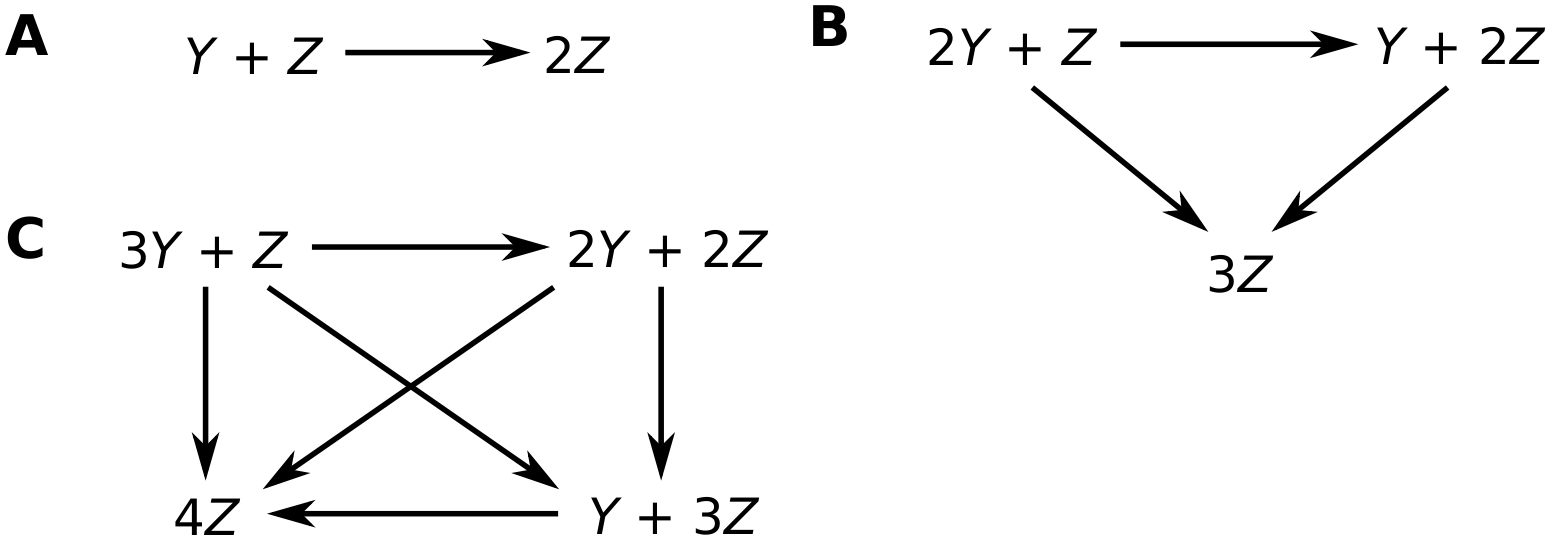}
 \caption{{\bf Schematic of cluster-stabilization reactions.} Examples of ligand-independent cluster-stabilization reactions involving unstable ($Y$) and stable ($Z$) open receptors of molecularities two (A), three (B), and four (C). Higher-order reactions follow the same pattern. Ligand-dependent reactions are identical except that FasL ($L$) must be added to each reacting state.}
 \label{fig:2}
\end{figure}

In our model, a cluster is indexed by a tuple $(L, X, Y, Z)$, where $L$ represents FasL and $X$, $Y$, and $Z$ are three posited forms of Fas, denoting closed, open and unstable, and open and stable, i.e., active and signaling, receptors, respectively. Within a cluster, we assumed a complete interaction graph and defined the reactions
\begin{subequations}
 \begin{align}
  X &\cee{<=>[k_{o}][k_{c}]} Y,\label{eq:cluster:reactions:open-close}\\
  Z &\cee{->[k_{u}]} Y,\label{eq:cluster:reactions:destabilize}\\
  jY + \left( i - j \right) Z &\cee{->[k_{s}^{\left( i \right)}]} \left( j - k \right) Y + \left( i - j + k \right) Z, &\quad
  \begin{cases}
   i = 2, \dots, m,\\
   j = 1, \dots, i,\\
   k = 1, \dots, j,
  \end{cases}
  \label{eq:cluster:reactions:ligand-independent}\\
  L + jY + \left( i - j \right) Z &\cee{->[k_{l}^{\left( i \right)}]} L + \left( j - k \right) Y + \left( i - j + k \right) Z, &\quad
  \begin{cases}
   i = 2, \dots, n,\\
   j = 1, \dots, i,\\
   k = 1, \dots, j.
  \end{cases}
  \label{eq:cluster:reactions:ligand-dependent}
 \end{align}
\end{subequations}
The first reaction describes spontaneous receptor opening and closing; the second, constitutive destabilization of open Fas; the third, ligand-independent receptor cluster-stabilization; and the fourth, ligand-dependent receptor cluster-stabilization (Figure \ref{fig:2}). The orders of the cluster-stabilization events are limited by the parameters $m$ and $n$, which capture the effects of receptor density and Fas coordination by FasL, respectively. Although only pair-stabilization ($m = n = 2$) has been observed experimentally \cite{scott:2009:nature}, higher-order analogues, for example, as facilitated by globular interactions, are not unreasonable.

Formally, these reactions are to be interpreted as state transitions on the space of cluster tuples. However, the reaction notation is suggestive, highlighting the contribution of each elementary event, which we modeled using constant reaction rates (for simplicity, we set uniform rate constants $k_{s}^{(i)}$ and $k_{l}^{(i)}$ for all ligand-independent and -dependent cluster-stabilization reactions of molecularity $i$, respectively). Then on making a continuum approximation, we reinterpreted the molecule numbers as local concentrations and applied the law of mass action to produce a dynamical system for each cluster in the concentrations $(l, x, y, z)$ of $(L, X, Y, Z)$. Validity of the model requires that the molecular concentrations are not too low and that the timescale of receptor conformational change is short compared to that of cluster dissociation.

To study the long-term behavior of the model, we solved the system at steady state (denoted by the subscript $\infty$). Introducing the nondimensionalizations
\begin{subequations}
 \begin{align}
  \xi &= \frac{x}{s},\\
  \eta &= \frac{y}{s},\\
  \zeta &= \frac{z}{s},\\
  \lambda &= \frac{l}{s},\\
  \tau &= k_{c} t,
 \end{align}
\end{subequations}
where $s$ is a characteristic concentration and $t$ is time, and
\begin{subequations}
 \begin{align}
  \kappa_{o} &= \frac{k_{o}}{k_{c}},\\
  \kappa_{u} &= \frac{k_{u}}{k_{c}},\\
  \kappa_{s}^{\left( i \right)} &= \frac{k_{s}^{\left( i \right)} s^{i - 1}}{k_{c}}, \quad i = 2, \dots, m,\\
  \kappa_{l}^{\left( i \right)} &= \frac{k_{l}^{\left( i \right)} s^{i}}{k_{c}}, \quad i = 2, \dots, n,
 \end{align}
\end{subequations}
this is
\begin{subequations}
 \begin{align}
  \xi_{\infty} &= \frac{\sigma - \zeta_{\infty}}{1 + \kappa_{o}},\\
  \eta_{\infty} &= \kappa_{o} \xi_{\infty},
 \end{align}
\end{subequations}
where $\sigma = \xi + \eta + \zeta$ is the nondimensional total receptor density, and $\zeta_{\infty}$ is given by considering
\begin{align}
 \frac{d \zeta}{d \tau} = \sum_{i = 2}^{m} \kappa_{s}^{\left( i \right)} \sum_{j = 1}^{i} \eta^{i} \zeta^{i - j} \sum_{k = 1}^{j} k + \lambda \sum_{i = 2}^{n} \kappa_{l}^{\left( i \right)} \sum_{j = 1}^{i} \eta^{i} \zeta^{i - j} \sum_{k = 1}^{j} k - \kappa_{u} \zeta
 \label{eq:cluster:zeta_tau}
\end{align}
and solving $d \zeta / d \tau = 0$ with $(\xi, \eta, \zeta) \mapsto (\xi_{\infty}, \eta_{\infty}, \zeta_{\infty})$, a polynomial in $\zeta_{\infty}$ of degree $\max \{ m, n \}$. Clearly, the model is bistable only if $\max \{ m, n \} \geq 3$ (two stable nodes must be separated by an unstable node as the model is effectively one-dimensional in $\zeta$).

We used $\zeta$ as a measure of the apoptotic activation of a cluster. In principle, all open receptors contribute to apoptotic signaling, but $\eta$ is small, at least at steady state (since $\kappa_{o} \ll 1$ due to the assumed prevalence of the closed form \cite{huang:1996:nature}), and so can be neglected.

\subsection*{Bistability and receptor clustering}
While $n$ measures the coordination capacity of FasL and hence may be equated with its oligomeric order (e.g., $n = 3$ in the biological context), an appropriate value for $m$, relating to the total receptor concentration, is somewhat more elusive. Therefore, we began our analysis by performing a simple receptor density estimate. Approximating the cell as a cube of linear dimension $\sim 10$ $\mu$m, the associated volume of $\sim 1$ pL implies the correspondence $1$ nM $\sim 600$ molecules $\sim 10^{-6}$ molecules/nm$^{2}$ on restricting to the membrane, i.e., by averaging over the surface area of $\sim 600$ $\mu$m$^{2}$. Thus, for a conservative receptor concentration estimate of $100$ nM \cite{bentele:2004:j-cell-biol,hua:2005:j-immunol,albeck:2008:mol-cell,albeck:2008:plos-biol}, the number of Fas molecules in the neighborhood of each receptor is only $\sim 1$, assuming a charateristic size of $100$ nm. We hence found that receptors may be very sparsely distributed. In this low density mode, high-order Fas interactions in the absence of ligand can be neglected ($m = 2$). Therefore, in this context, bistability is possible only if $n \geq 3$, and the trimerism of FasL thus demonstrates the lowest-order complexity required for bistability.

From the form of $d \zeta / d \tau$, this bistability is reversible as a function of the FasL concentration $\lambda$ since the governing polynomial for $\zeta_{\infty}$ is of degree only $m = 2$ at $\lambda = 0$. This suggests that at the cluster level, the cell death decision can be reversed, which may have adverse effects on cellular and genomic integrity. However, irreversible bistability at higher receptor densities may also be achieved. Researchers have observed tendencies for death receptors both to pre-associate as dimers or trimers \cite{chan:2000:science,siegel:2000:science,chan:2007:cytokine} and to selectively localize onto membrane lipid rafts \cite{muppidi:2004:nat-immunol}. The result of either of these processes may be to increase the local receptor concentration. In this high density mode, we set $m \geq 3$, as the preceeding approximation is no longer valid. Irreversible bistability then becomes attainable, representing a committed cell death decision.

For the remainder of the study, we incorporated both the low and high receptor density regimes into a single model by setting $m = 3$, using $\sigma$ as a continuous transition parameter. Furthermore, we set $n = 3$ to correspond to observed biology.

\subsection*{Characterization of the steady-state surface}
\begin{figure}[!ht]
 \includegraphics{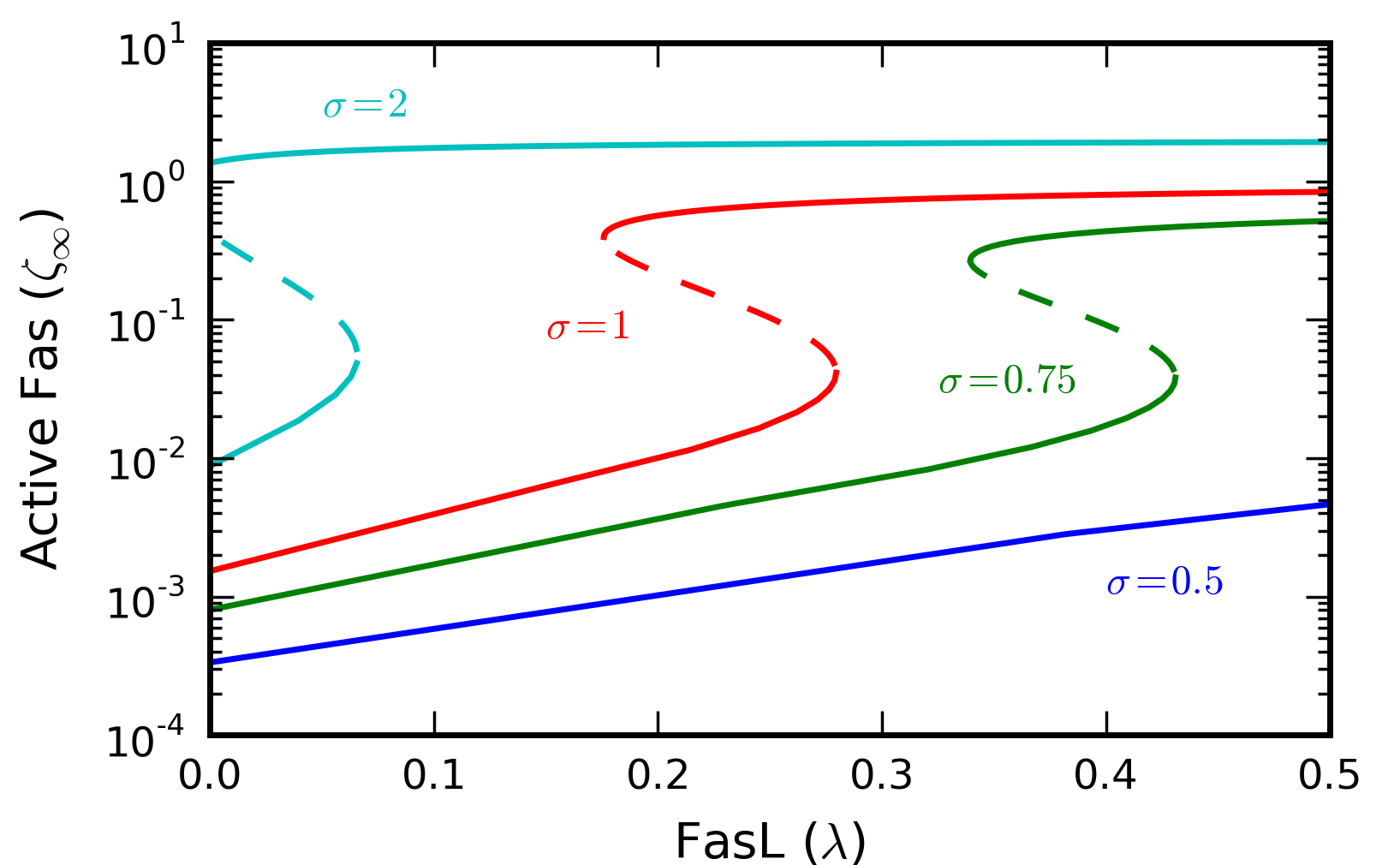}
 \caption{{\bf Steady-state activation curves.} The steady-state active Fas concentration $\zeta_{\infty}$ shows bistability and hysteresis as a function of the FasL concentration $\lambda$ (stable, solid lines; unstable, dashed lines). At low receptor concentrations $\sigma$, the bistability is reversible, but irreversibility emerges for $\sigma$ sufficiently high, representing a committed cell death decision. All parameters set at baseline values unless otherwise noted.}
 \label{fig:3}
\end{figure}

Calculation of the steady-state activation curves showed that the model indeed exhibits bistability (Figure \ref{fig:3}) for reasonable parameter choices (Methods). Thus, we established the possiblity of a novel bistability mechanism in extrinsic apoptosis. The associated hysteresis enables threshold switching between well-separated low and high activation states. Biologically, these define local signals of life and death, which are integrated at the cell level to compute the overall apoptotic response.

\begin{figure}[!ht]
 \includegraphics{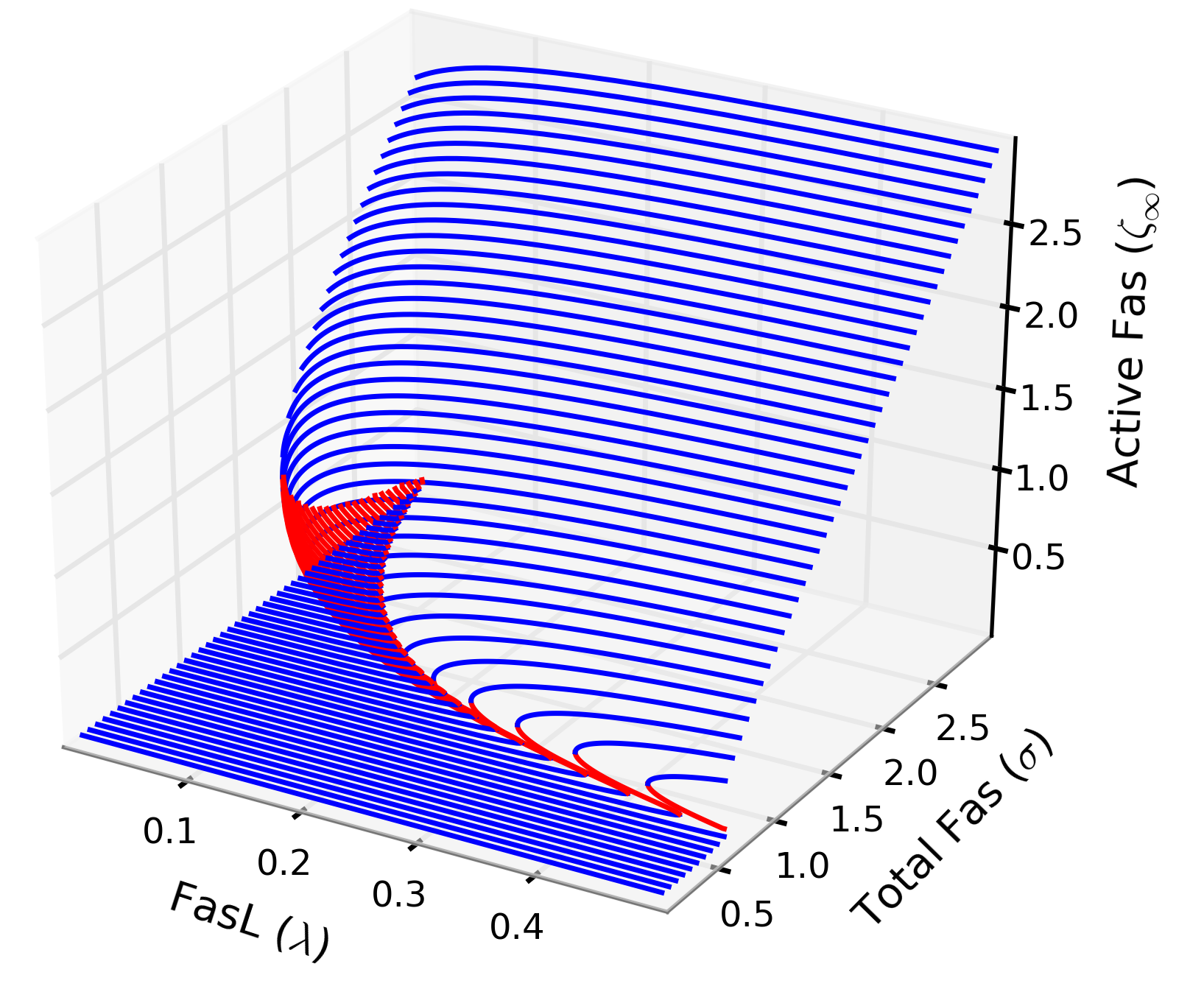}
 \caption{{\bf Steady-state activation surface.} The steady-state surface for the active Fas concentration $\zeta_{\infty}$ as a function of the FasL and total Fas concentrations $\lambda$ and $\sigma$, respectively, is folded, indicating the existence of singularities, across which the system's steady-state behavior switches between monostability and bistability (stable, blue; unstable, red). All parameters set at baseline values unless otherwise noted.}
 \label{fig:4}
\end{figure}

\begin{figure}[!ht]
 \includegraphics{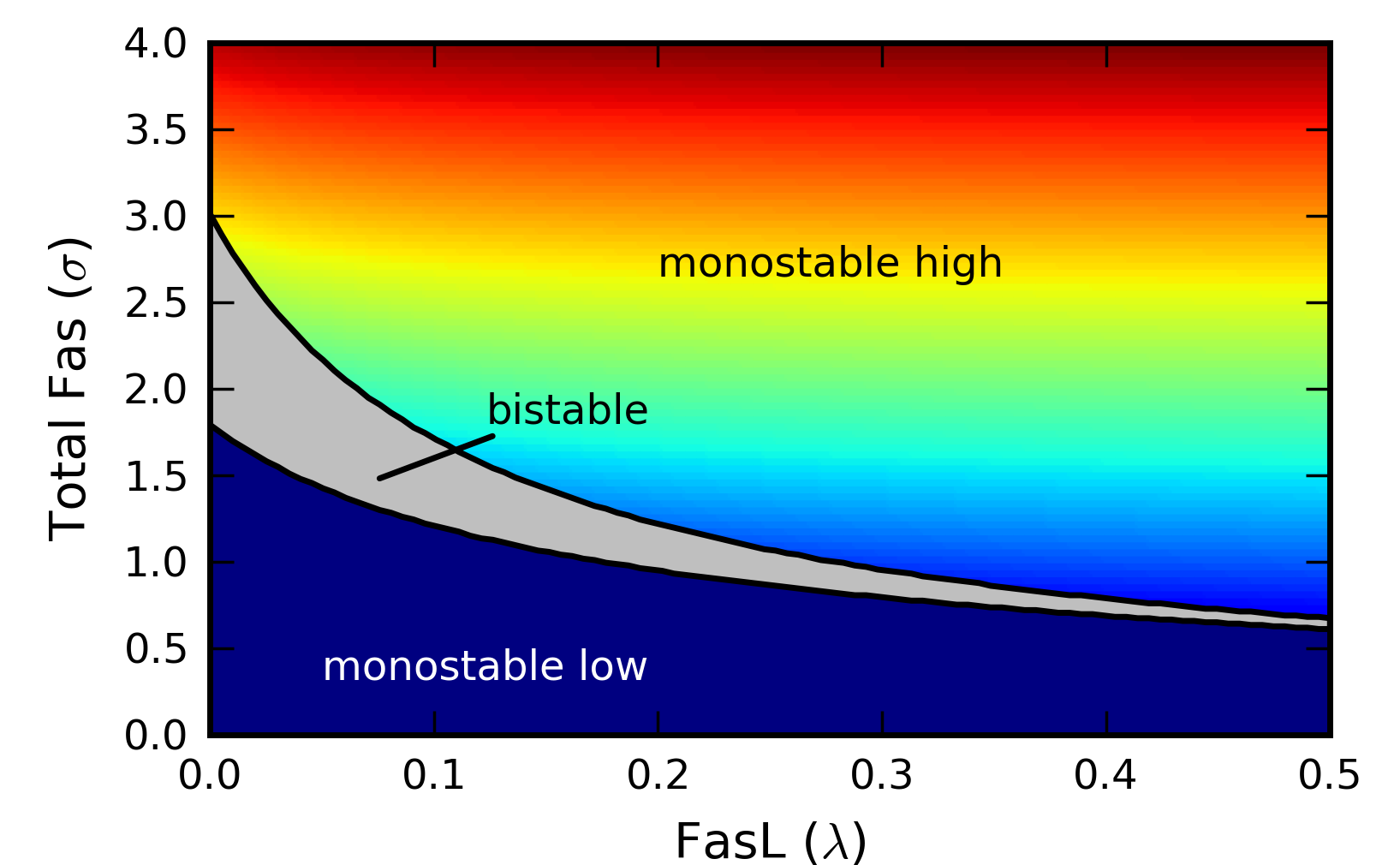}
 \caption{{\bf Steady state diagram.} Steady state diagram identifying the regions of parameter space supporting monostability (colored) or bistability (gray) as a function of the FasL and total Fas concentrations $\lambda$ and $\sigma$, respectively. The monostable region is colored as a heat map corresponding to the steady-state active Fas concentration $\zeta_{\infty}$. Irreversible bistability is indicated by the extension of the bistable region to the axis $\lambda = 0$.}
 \label{fig:5}
\end{figure}

As per the previous analysis, reversibility of the bistability is dependent on $\sigma$, with irreversibility emerging for $\sigma$ sufficiently high. This suggests a bivariate parameterization of $\zeta_{\infty}$, producing a multivalued steady-state surface over $(\lambda, \sigma)$-space (Figure \ref{fig:4}). The result is a cusp catastrophe, an elementary object of catastrophe theory, which studies how small perturbations in certain parameters can lead to large and sudden changes in the behavior of a nonlinear system \cite{arnold:1992:springer-verlag}. A more instructive view of the dependence of the model's qualitative structure on $\lambda$ and $\sigma$ is shown in Figure \ref{fig:5}.

\subsection*{Sensitivity and robustness analyses}
\begin{figure}[!ht]
 \includegraphics{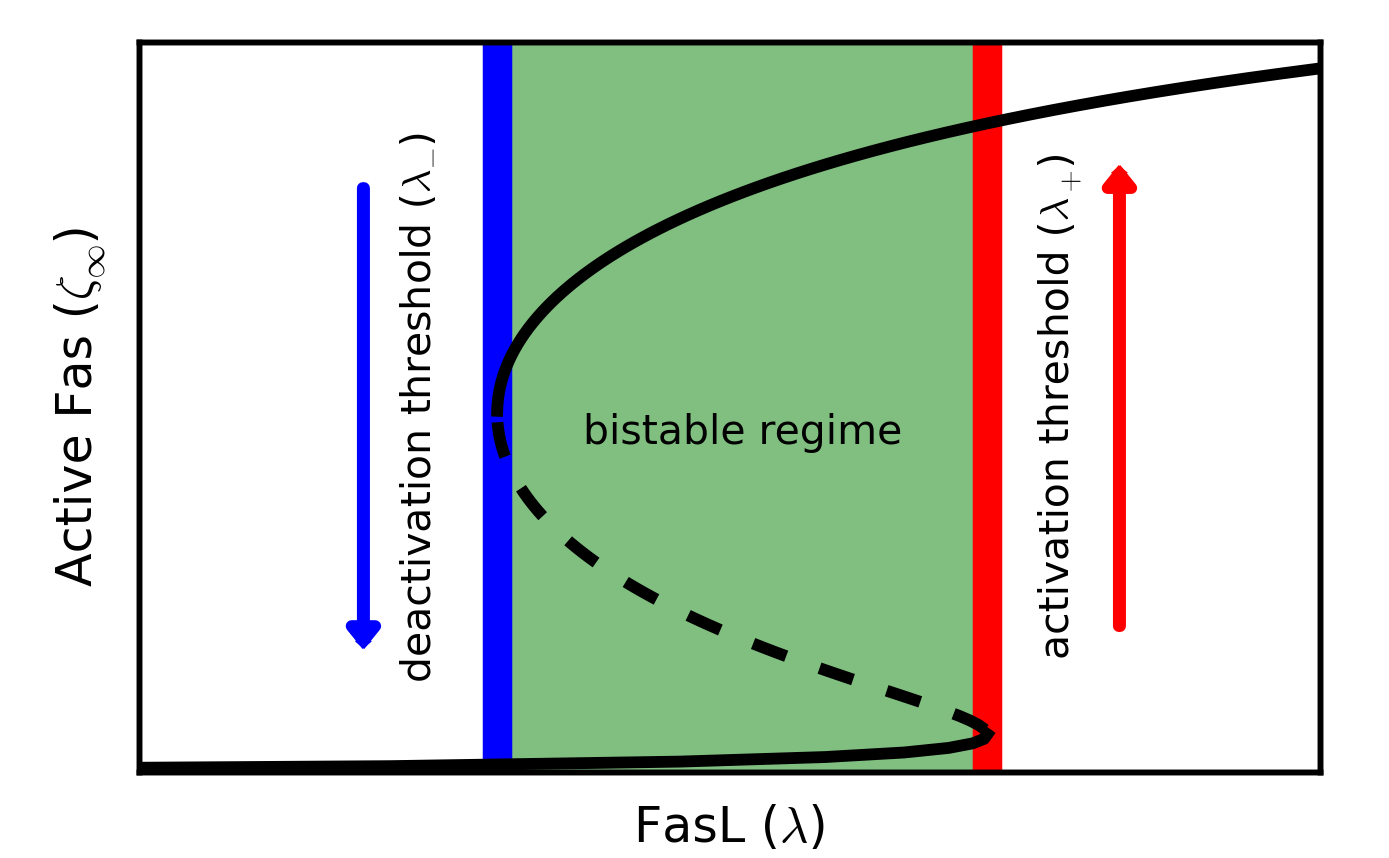}
 \caption{{\bf Bistability thresholds.} The activation (red) and deactivation (blue) thresholds $\lambda_{\pm}$ characterizing the bistable regime (green) are defined as the concentrations $\lambda$ of FasL at which the steady-state active Fas concentration $\zeta_{\infty}$ (black) switches discontinuously from one branch to the other (stable, solid line; unstable, dashed line).}
 \label{fig:6}
\end{figure}

We then focused on the activation and deactivation thresholds $\lambda_{\pm}$, respectively, defining the bistable regime. These are the points at which the steady state switches discontinuously from one branch to the other, and are given by the values of $\lambda$ at which the hysteresis curve turns, i.e., at $\partial \lambda / \partial \zeta_{\infty} = 0$ (Figure \ref{fig:6}). We performed a sensitivity analysis of $\lambda_{\pm}$ by measuring the effects of perturbing the model parameters about baseline values (Methods). For each threshold-parameter pair, we computed a normalized sensitivity $\Sigma$ by linear regression.

\begin{figure}[!ht]
 \includegraphics{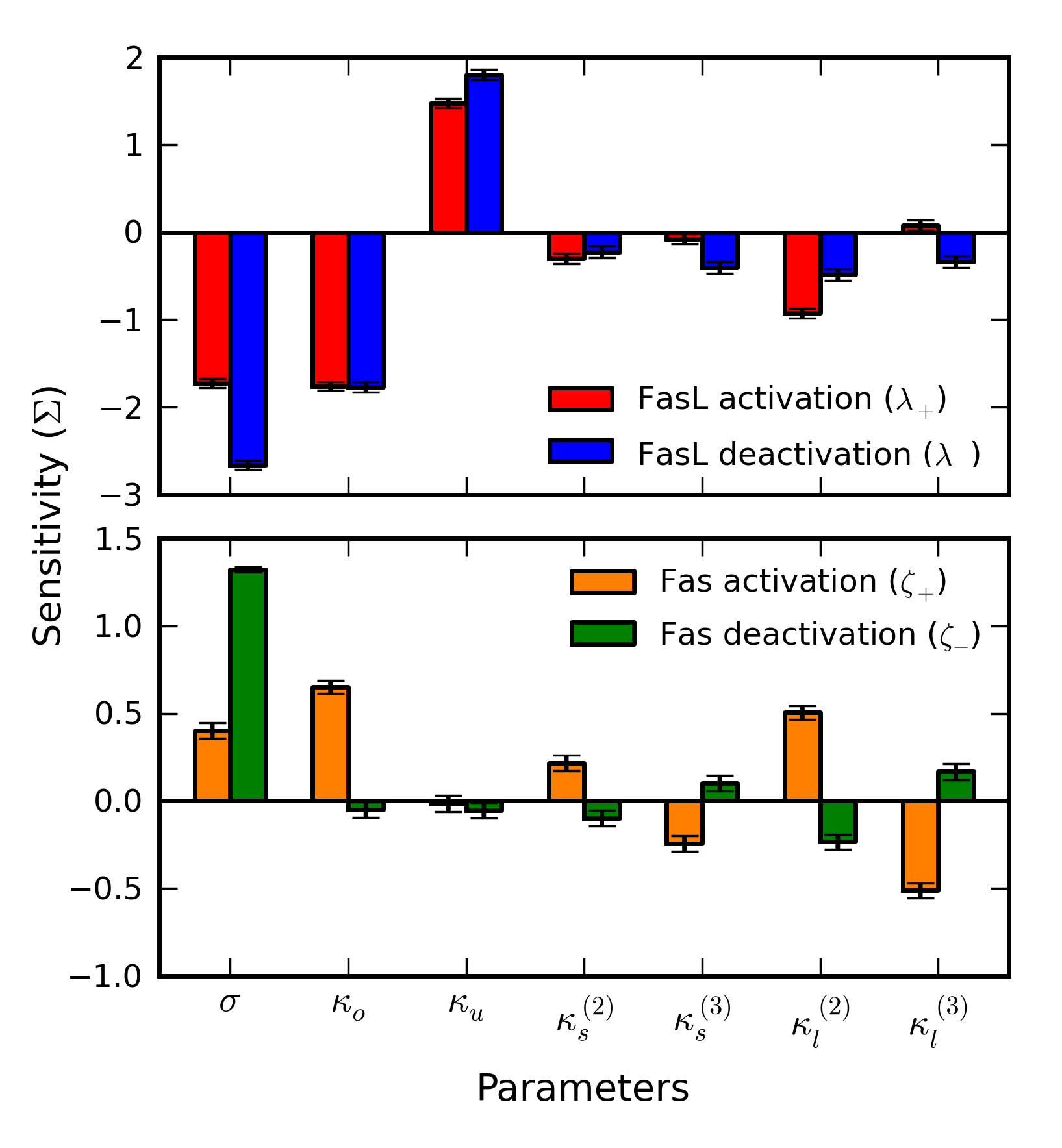}
 \caption{{\bf Sensitivity analysis of bistability thresholds.} The robustness of the bistability thresholds is investigated by measuring the effects of perturbating the model parameters about baseline values. For each threshold-parameter pair, a normalized sensitivity $\Sigma$ is computed by linear regression. Top, sensitivities for the FasL thresholds $\lambda_{\pm}$; bottom, sensitivities for the corresponding Fas thresholds $\zeta_{\pm}$ at FasL concentrations $\lambda = \lambda_{\pm}$, respectively.}
 \label{fig:7}
\end{figure}

Strong effects of $\sigma$, $\kappa_{o}$, and $\kappa_{u}$ were observed (Figure \ref{fig:7}); for the corresponding Fas thresholds $\zeta_{\pm}$ at $\lambda = \lambda_{\pm}$, respectively, the parameters $\sigma$, $\kappa_{o}$, $\kappa_{l}^{(2)}$, and $\kappa_{l}^{(3)}$ were emphasized. Thus, the bistability thresholds do not appear particularly robust. However, the data reveal that essentially all parameter sets sampled were bistable. This suggests a weaker form of robustness, namely, robustness of bistability, which nevertheless supports life and death decisions over a wide operating range.

\begin{figure}[!ht]
 \includegraphics{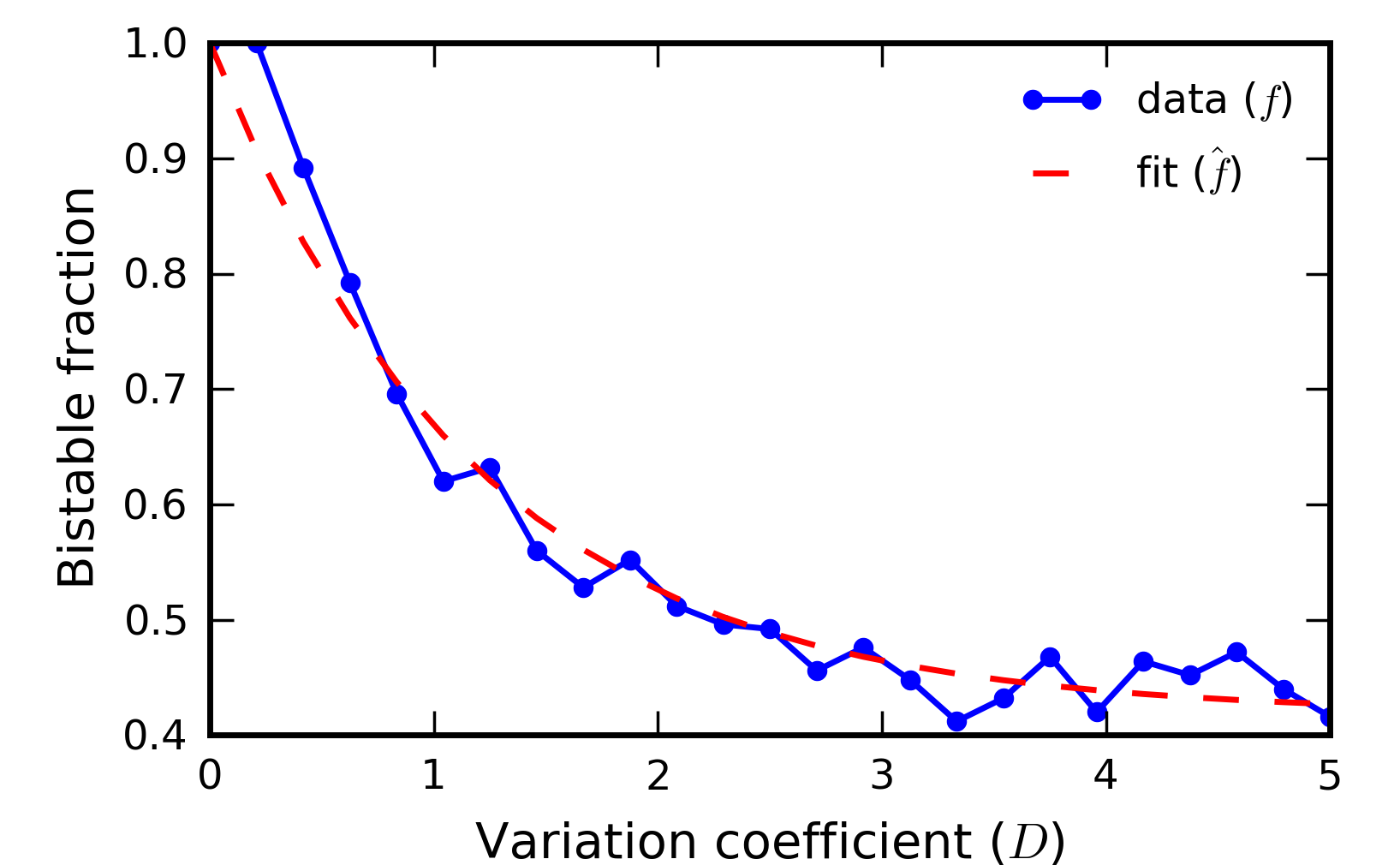}
 \caption{{\bf Robustness of bistability.} The fraction $f$ of parameter sets that exhibit bistability as a function of the sampling variability $D$ follows the exponential form $\hat{f} = f_{\infty} + (1 - f_{\infty}) e^{-D / D_{0}}$, where $f_{\infty}$ is the asymptotic bistable fraction. The fitted value of $f_{\infty} \approx 0.4$ suggests that this robustness remains substantial even as $D \to \infty$.}
 \label{fig:8}
\end{figure}

To probe this further, we sampled parameters with increasing spread $D$ about baseline values and computed the fraction $f$ of parameter sets that remained bistable (Methods). The results show that $f$ has an exponential form (Figure \ref{fig:8}). Extrapolating to $D \to \infty$, the data suggest an asymptotic bistable fraction of $f_{\infty} \approx 0.4$. Hence, robustness of bistability remains substantial even under significant parameter variation.

\subsection*{Cell-level cluster integration}
Thus far, we have considered only the apoptotic activation of an individual cluster. To obtain the more biologically relevant cell-level activation, we must integrate over all clusters. In principle, this integration should account for intercluster transport as well as any intrinsic differences between clusters, e.g., as due to spatial inhomogeneities. Here, however, we provide as demonstration only a very simple integration scheme. Specifically, we assumed that clusters are identical (apart from their parameter values, which are drawn randomly) and independent, and that FasL is homogeneous over the cell membrane. Then we can express the normalized cell activation as
\begin{align}
 \zeta_{\infty}^{\cell} \left( \lambda \right) = \frac{\sum_{i} \zeta_{\infty, i} \left( \lambda \right)}{\sum_{i} \sigma_{i}},
\end{align}
where the subscript $i$ denotes reference to cluster $i$.

\begin{figure}[!ht]
 \includegraphics{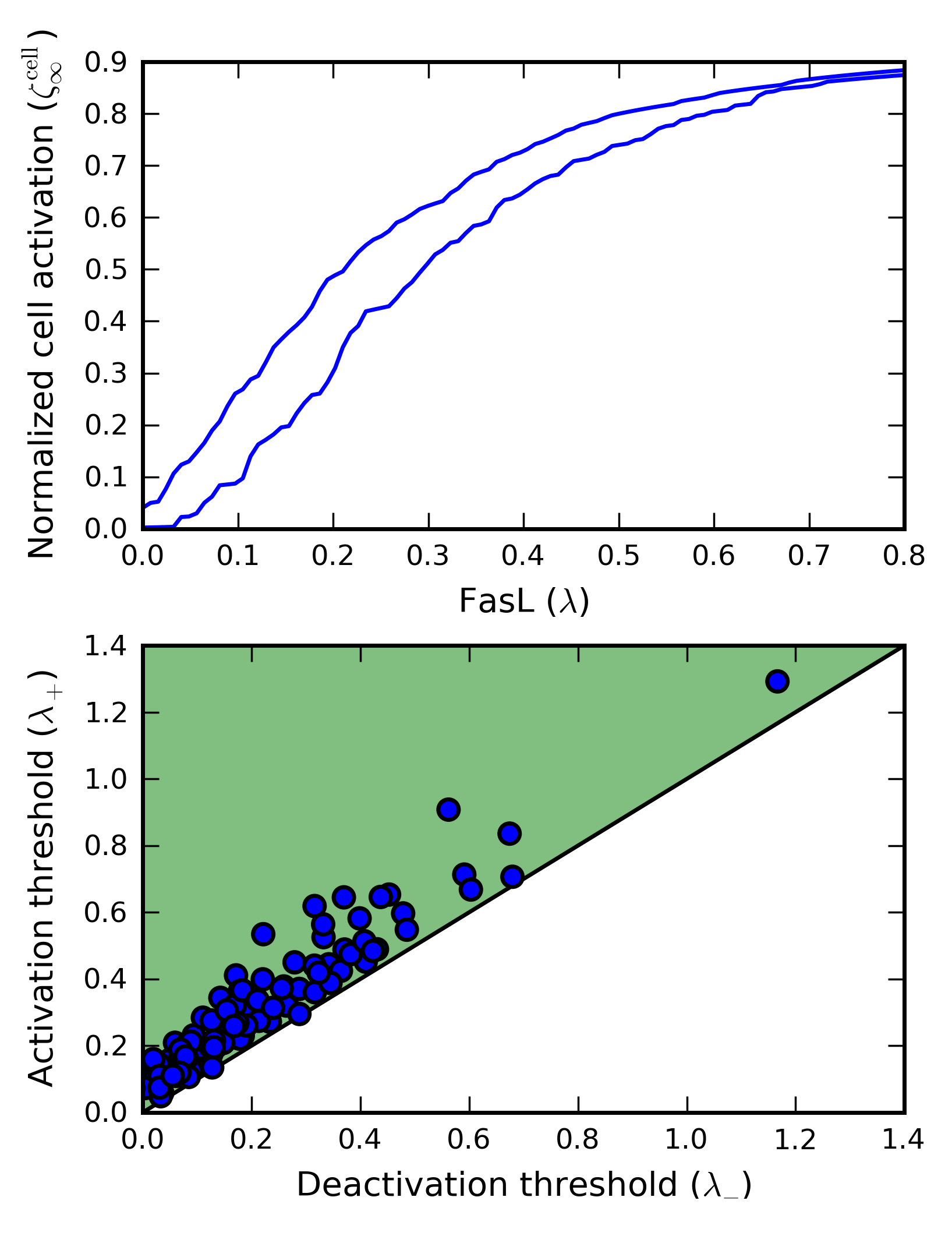}
 \caption{{\bf Cell-level cluster integration.} The apoptotic signals of all Fas clusters are integrated to produce a normalized cell activation $0 \leq \zeta_{\infty}^{\cell} \leq 1$. The resulting hysteresis curve on $\zeta_{\infty}^{\cell}$ as a function of the FasL concentration $\lambda$ is graded due to the heterogeneity of the bistability thresholds $\lambda_{\pm}$ across the clusters (top). Despite this variability, a strong linear dependence persists between $\lambda_{\pm}$ (bottom; the valid region $\lambda_{+} > \lambda_{-}$ is shown in green).}
 \label{fig:9}
\end{figure}

A characteristic cell-level hysteresis curve is shown in Figure \ref{fig:9}. As is immediately evident, such integration is a smoothing operator, averaging over the sharp thresholds of each cluster. Thus, the cell-level signal may be graded even though its constituents are not. Note, however, that the lack of a sudden switch from low to high Fas signaling does not necessarily imply the same at the level of the caspases which ultimately govern cell death, as downstream components may possess switching behaviors \cite{bentele:2004:j-cell-biol,eissing:2004:j-biol-chem,bagci:2006:biophys-j,legewie:2006:plos-comput-biol,cui:2008:plos-one}.

\subsection*{Model discrimination}
Finally, we sought to outline protocols to experimentally discriminate our model against the prevailing crosslinking model \cite{lai:2004:math-biosci-eng}, which we considered in a slightly simplified form \cite{harrington:2008:theor-biol-med-model}. To be precise, the crosslinking model that we used has the reactions
\begin{subequations}
 \begin{align}
  L + R &\cee{<=>[{3 k_{+}}][k_{-}]} C_{1},\\
  C_{1} + R &\cee{<=>[{2 k_{+}}][{2 k_{-}}]} C_{2},\\
  C_{2} + R &\cee{<=>[k_{+}][{3 k_{-}}]} C_{3},
 \end{align}
\end{subequations}
where $L$ is FasL, $R$ is Fas, and $C_{i}$ is the complex FasL:Fas$_{i}$ for $i = 1, 2, 3$. With
\begin{subequations}
 \begin{align}
  \lambda &= \frac{l}{s},\\
  \rho &= \frac{r}{s},\\
  \gamma_{i} &= \frac{c_{i}}{s}, \quad i = 1, 2, 3,\\
  \kappa &= \frac{k_{-}}{k_{+} s},\\
  \tau &= k_{+} s t,
 \end{align}
\end{subequations}
(continuing the notational convention that lowercase letters denote the concentrations of their uppercase counterparts), the steady-state solution under mass-action dynamics is
\begin{subequations}
 \begin{align}
  \gamma_{1, \infty} &= 3 \lambda_{\infty} \left( \frac{\rho_{\infty}}{\kappa} \right),\\
  \gamma_{2, \infty} &= 3 \lambda_{\infty} \left( \frac{\rho_{\infty}}{\kappa} \right)^{2},\\
  \gamma_{3, \infty} &= \lambda_{\infty} \left( \frac{\rho_{\infty}}{\kappa} \right)^{3},
 \end{align}
\end{subequations}
where
\begin{subequations}
 \begin{align}
  \lambda_{\infty} &= \frac{\Lambda}{1 + 3 \left( \rho_{\infty} / \kappa \right) + 3 \left( \rho_{\infty} / \kappa \right)^{2} + \left( \rho_{\infty} / \kappa \right)^{3}},\\
  \rho_{\infty} &= \frac{1}{2} \left[ \sqrt{\left( 3 \Lambda + \kappa - \sigma \right)^{2} + 4 \kappa \sigma} - \left( 3 \Lambda + \kappa - \sigma \right) \right],
 \end{align}
\end{subequations}
and
\begin{subequations}
 \begin{align}
  \Lambda &= \lambda + \gamma_{1} + \gamma_{2} + \gamma_{3},\\
  \sigma &= \rho + \gamma_{1} + 2 \gamma_{2} + 3 \gamma_{3}
 \end{align}
\end{subequations}
are the total ligand and receptor concentrations, respectively. In analogy with our proposed model, hereafter called the cluster model, we used
\begin{align}
 \zeta \equiv \gamma_{1} + 2 \gamma_{2} + 3 \gamma_{3} = \sigma - \rho
\end{align}
as a measure of the apoptotic signal.

\subsubsection*{Hyperactive mutants}
Clearly, the crosslinking model has only one steady state, while the cluster model is capable of bistability. This hence provides a ready discrimination criterion. Although tracing out the associated hysteresis curve may be problematic, we can nevertheless probe for bistability by using hyperactive mutants, e.g., the mutation of Ile 313 to Asp in Fas, which stabilizes the open conformation and enhances apoptotic activity \cite{scott:2009:nature}.

Specifically, we considered an experimental setup in which the concentrations of FasL and Fas, both wildtype and mutant, can be controlled, and in which the apoptotic signal can be measured, e.g., through the degree of FADD binding or of caspase activation. Hence we can map out the response curves at various levels of mutant penetrance. Denoting mutant Fas by $Z_{\Delta}$ (we assumed that mutant Fas cannot close, so there is no distinction between the stable and unstable open forms), we characterized the mutant penetrance by the mutant population fraction $\Delta = \zeta_{\Delta} / \overline{\sigma}$, where $\zeta_{\Delta} = z_{\Delta} / s$ is the nondimensional mutant Fas concentration and $\overline{\sigma} = \sigma + \zeta_{\Delta}$ is the total receptor concentration, composed of contributions from both wildtype ($\sigma$) and mutant ($\zeta_{\Delta}$) forms. We assumed no other functional differences between wildtype and mutant Fas.

Proceeding first for the crosslinking model, at fixed $\overline{\sigma}$, the amount of Fas bound by FasL is determined only by $\Lambda$. Hence we assumed that a fraction $\varphi = \varphi (\Lambda)$ of all receptors is bound. Since wildtype and mutant Fas are functionally identical by assumption, the fraction bound for each of the wildtype and mutant populations is also $\varphi (\Lambda)$ by independence of the recruitment process. Therefore, under the crosslinking model, varying $\Delta$ at fixed $\overline{\sigma}$ yields an invariant response curve for the active wildtype Fas fraction $\varphi = \zeta_{\infty} / \sigma$.

In contrast, for the cluster model, we expected mutant receptor cluster-interactions to affect the wildtype response. Accordingly, the reactions \eqref{eq:cluster:reactions:ligand-independent} and \eqref{eq:cluster:reactions:ligand-dependent} were amended for interaction with $Z_{\Delta}$ by replacing with
\begin{subequations}
 \begin{align}
  jY + kZ + \left( i - j - k \right) Z_{\Delta} &\cee{->[k_{s}^{\left( i \right)}]} \left( j - k' \right) Y + \left( k + k' \right) Z + \left( i - j - k \right) Z_{\Delta}, &\quad
  \begin{cases}
   i = 2, \dots, m,\\
   j = 1, \dots, i,\\
   k = 0, \dots, i - j,\\
   k' = 1, \dots, j,
  \end{cases}\\
  L + jY + kZ + \left( i - j - k \right) Z_{\Delta} &\cee{->[k_{l}^{\left( i \right)}]} L + \left( j - k' \right) Y + \left( k + k' \right) Z + \left( i - j - k \right) Z_{\Delta}, &\quad
  \begin{cases}
   i = 2, \dots, n,\\
   j = 1, \dots, i,\\
   k = 0, \dots, i - j,\\
   k' = 1, \dots, j,
  \end{cases}
 \end{align}
\end{subequations}
respectively. This gives the analogue
\begin{align}
 \frac{d \zeta}{d \tau} = \sum_{i = 2}^{m} \kappa_{s}^{\left( i \right)} \sum_{j = 1}^{i} \eta^{j} \sum_{k = 0}^{i - j} \zeta^{k} \zeta_{\Delta}^{i - j - k} \sum_{k' = 1}^{j} k' + \lambda \sum_{i = 2}^{n} \kappa_{l}^{\left( i \right)} \sum_{j = 1}^{i} \eta^{j} \sum_{k = 0}^{i - j} \zeta^{k} \zeta_{\Delta}^{i - j - k} \sum_{k' = 1}^{j} k' - \kappa_{u} \zeta
\end{align}
of \eqref{eq:cluster:zeta_tau}.

\begin{figure}[!ht]
 \includegraphics{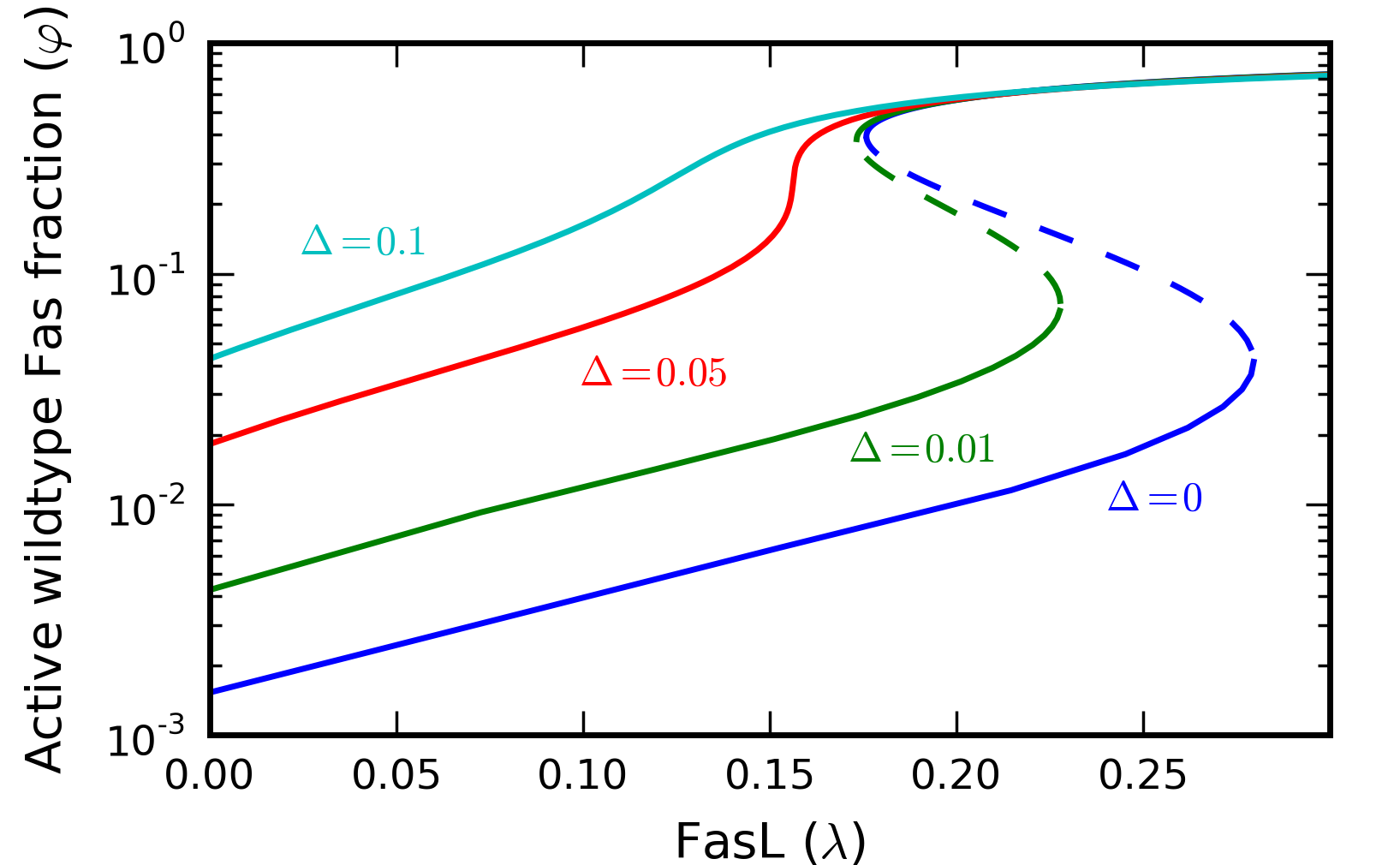}
 \caption{{\bf Model discrimination using hyperactive mutants.} The wildtype response curve, giving the steady-state active wildtype Fas fraction $\varphi$ as a function of the FasL concentration $\lambda$ (stable, solid lines; unstable, dashed lines), of the cluster model varies with the mutant population fraction $\Delta$, reflecting receptor interactions absent in the crosslinking model. The total receptor concentration is fixed at $\overline{\sigma} = 1$. All parameters set at baseline values unless otherwise noted.}
 \label{fig:10}
\end{figure}

As seen in Figure \ref{fig:10}, receptor interactions indeed cause the apoptotic signal to increase with $\Delta$ even after accounting for the effect of mutants. This is because mutants can activate wildtype receptors by pushing the cluster past its switching threshold. Furthermore, the convergence to the active cluster state at high $\lambda$ provides evidence for bistability. Thus, the variance of the $\varphi$-response curve at various $\Delta$ can be used for model discrimination.

\subsubsection*{Steady-state invariants}
Alternatively, if working with mutants should prove difficult, we provide also a discrimination test based on steady-state invariants, i.e., functions that vanish at steady state. Clearly, for each model, $\dot{\zeta} \equiv d \zeta / d \tau$ provides a steady-state invariant since $\dot{\zeta} = 0$ necessarily at steady state. However, the difficulty lies in expressing $\dot{\zeta}$ solely in terms of variables that are experimentally accessible. For example, current technology may not allow the concentrations $\gamma_{i}$ to be measured accurately, if at all. Therefore, all such variables must be eliminated. Rate constants were considered parameters and so were not subject to this rule.

We assumed the same experimental setup as above and hence expressed each model invariant in terms of $\Lambda$, $\sigma$, and $\zeta_{\infty}$, giving functions of the form $\dot{\zeta} (\Lambda, \sigma, \zeta_{\infty}; \boldsymbol{a})$, where $\boldsymbol{a}$ encompasses all model parameters (Methods). The task then is to use $|\dot{\zeta}|$, with $(\Lambda, \sigma, \zeta_{\infty})$ provided by experiment, to assess the fit of a model. However, the parameters $\boldsymbol{a}$ remain unknown, so this assessment cannot proceed directly. Instead, we considered the best possible fit $\min_{\boldsymbol{a}} |\dot{\zeta}|$ over all parameters. A high value of $\min_{\boldsymbol{a}} |\dot{\zeta}|$ indicates a poor best-case fit and hence that a model is unlikely to be correct. Clearly, prior knowledge of $\boldsymbol{a}$ can be used to guide the invariant to biologically plausible fits.

\begin{figure}[!ht]
 \includegraphics{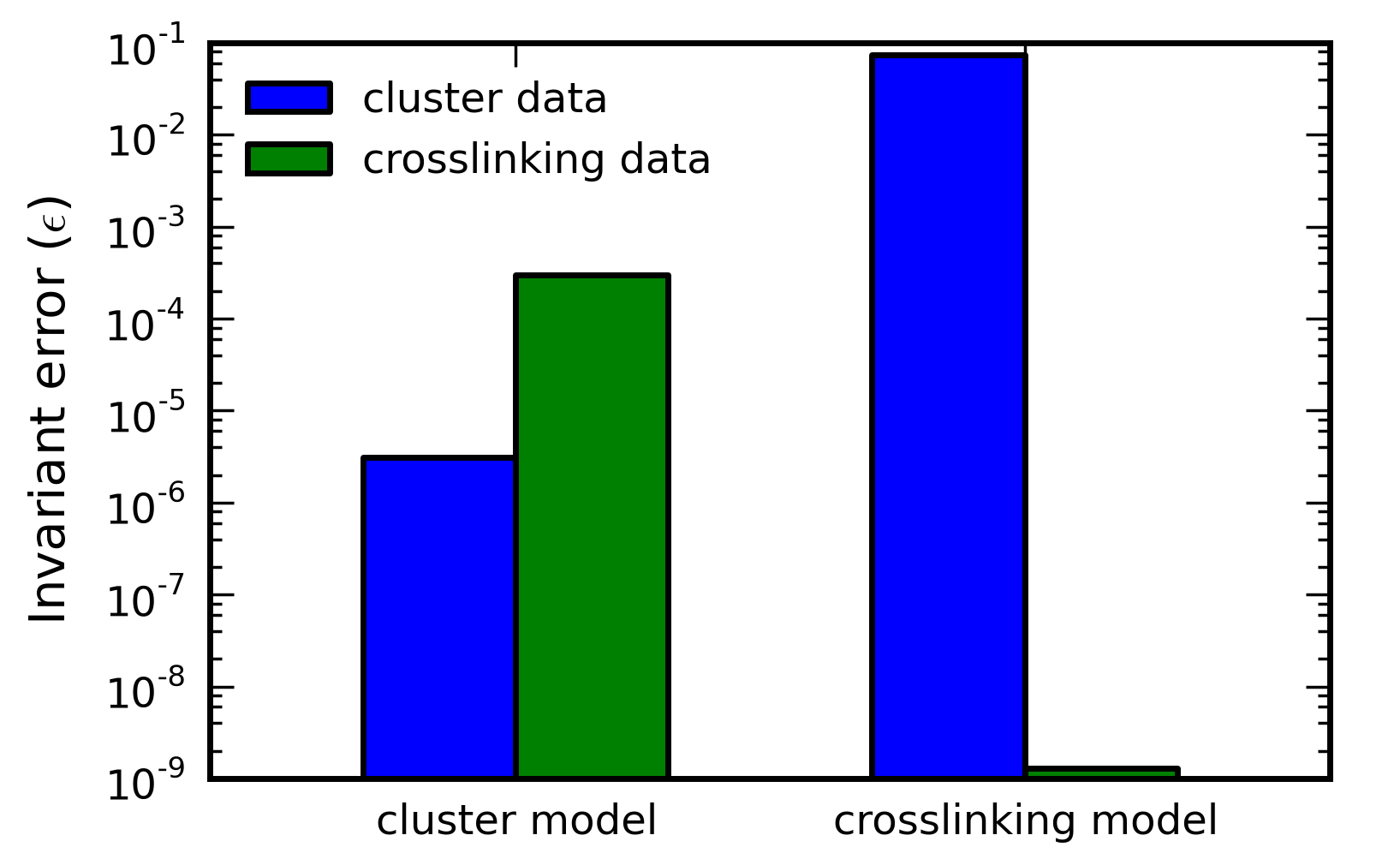}
 \caption{{\bf Model discrimination using steady-state invariants.} Steady-state invariants are fit to synthetic data generated from each model. For each model-data pair, the invariant error $\epsilon$ is minimized over the model parameter space. The results suggest that invariant minimization can correctly identify the model from the data.}
 \label{fig:11}
\end{figure}

To demonstrate that model discrimination using steady-state invariants is practical, we generated synthetic data from each model, calculating the accessible concentrations $(\Lambda, \sigma, \zeta_{\infty})$ for each parameter set. This gives two sets of model-generated data. For each data set, we computed the best-fit invariant error $\epsilon = \min_{\boldsymbol{a}} \rms |\dot{\zeta}|$ for each model, where $\rms$ is the root mean square operator. The results suggest that this test can correctly identify the model from the data (Figure \ref{fig:11}).

The systems that we have presently considered are simple enough that experimentally inaccessible variables can be eliminated by hand. For more complicated systems, the tools of computational algebraic geometry, notably Gr\"{o}bner bases, may prove useful; for such an application, see \cite{manrai:2008:biophys-j}.

\section*{Discussion}
In this work, we showed through analysis of a mathematical model that receptor clustering can support bistability and hysteresis in apoptosis through a higher-order analogue of biologically observed Fas pair-stabilization \cite{scott:2009:nature}. Hence we add to the signal processing activities in which receptor clustering has been suggested to participate \cite{bray:1998:nature,sourjik:2004:trends-microbiol,endres:2008:mol-syst-biol}. This bistability plays an important functional role by enabling robust threshold switching between life and death states. Significantly, our results indicate potential key roles for ligand trimerism \cite{ashkenazi:1998:science} and receptor pre-association \cite{chan:2000:science,siegel:2000:science,chan:2007:cytokine} and localization onto membrane lipid rafts \cite{muppidi:2004:nat-immunol}. Thus, we provide novel interpretations for these phenomena within the unified context of bistability.

Our model suggests an additional cell death decision, supplementing those that have been studied previously \cite{bentele:2004:j-cell-biol,eissing:2004:j-biol-chem,bagci:2006:biophys-j,legewie:2006:plos-comput-biol,cui:2008:plos-one}. Critically, the proposed decision is implemented upstream at the very death receptors that initially detect the death signal encoded by FasL. This decision is therefore apical in that it precedes all others in the system. Consequently, it operates independently of all intracellular components and so offers a general mechanism for bistability, even in cell lines with, for example, only feedforward caspase-activation networks \cite{bentele:2004:j-cell-biol,hua:2005:j-immunol,albeck:2008:plos-biol,okazaki:2008:j-theor-biol}. Thus, receptor cluster-activation may explain how an effective apoptotic decision is implemented in such cells. Moreover, this suggests a novel target for induced cell termination in the treatment of disease \cite{thompson:1995:science}.

We believe that our model provides an attractive theory for the observed biology. Although unlikely to be correct in mechanistic detail, the model may nevertheless reflect reality at a qualitative level. The significance of our work hence lies in its capacity to guide future research. We therefore readily invite experiment, which can reveal the true nature of the molecular mechanisms involved. Given their structural and functional homology, similar investigations on other members of the tumor necrosis factor receptor family may also prove fruitful. Such work serves to further our understanding of the formation and mode of action of complex signaling platforms such as the DISC, which in this view may be considered the macromolecular aggregates of active Fas.

\section*{Methods}

\subsection*{Parameter selection}
The rationale for the choices $m = n = 3$ is presented in the text; here, we further defend these by noting that no new behaviors are introduced with $m$ or $n > 3$. The remaining parameter values were guided by the following considerations. Specifically, we required $\kappa_{o}$ and $\kappa_{u} \ll 1$ due to the assumed stabilities of the receptor species; all other parameters were assumed to be close to $\bigO (1)$. Within these constraints, parameters were selected to ensure that $\lambda_{+}$ is of the correct order of magnitude \cite{bentele:2004:j-cell-biol,hua:2005:j-immunol,albeck:2008:mol-cell,albeck:2008:plos-biol}. The baseline parameter values used were $\sigma = 1$, $\kappa_{o} = 2 \times 10^{-3}$, $\kappa_{u} = 10^{-3}$, $\kappa_{s}^{(2)} = 0.1$, $\kappa_{s}^{(3)} = 0.5$, $\kappa_{l}^{(2)} = 1$, and $\kappa_{l}^{(3)} = 5$.

\subsection*{Parameter sampling}
To analyze the effects of variability in the model parameters, parameter values were sampled from a log-normal distribution, characterized by a variation coefficient $D$, defined as the ratio of the standard deviation to the median of the distribution. For the sensitivity analysis, $500$ parameter sets were drawn at $D = 0.25$; for the robustness analysis, $250$ parameter sets were drawn over $0 \leq D \leq 5$; and for the cell-level integration, $100$ parameter sets were drawn at $D = 0.25$. All parameters were drawn about baseline median values.

\subsection*{Sensitivity analysis}
For each threshold-parameter pair, linear regression was performed on the threshold data against the parameter data, each normalized by reference values. For parameters, the reference is the baseline (median) value; for thresholds, the reference is the threshold computed at baseline parameters. The normalized sensitivity $\Sigma$ was defined as the slope of the linear regression.

\subsection*{Steady-state invariants}
The cluster invariant was derived by considering $\dot{\zeta}$ at steady state, i.e., with $(\xi, \eta) \mapsto (\xi_{\infty}, \eta_{\infty})$, and identifying $\lambda \mapsto \Lambda$. Similarly, the crosslinking invariant was derived by considering $\dot{\zeta}$ at steady state and identifying $\rho_{\infty} \mapsto \sigma - \zeta_{\infty}$. For the full forms of the invariants, see Protocol S1.

For the model discrimination computation, $100$ parameter sets $(\Lambda, \sigma)$ were drawn from log-normal distributions with median $(0.2, 1)$ at $D = (0.2, 0.5)$. The active Fas concentration $\zeta_{\infty}$ was calculated for each parameter set for each model at baseline parameters ($\kappa = 0.1$ for the crosslinking model); for the cluster model, if bistability was observed, one of the stable values of $\zeta_{\infty}$ was chosen at random. The invariant error $\epsilon$ was minimized using SLSQP with a lower bound of $10^{-3}$ for all parameters.

\subsection*{Computational platform}
All calculations were performed with Sage 4.5 \cite{stein:2008:proc-int-symp-symb-algebra-comput}, using NumPy/SciPy \cite{oliphant:2007:comput-sci-eng} for numerical computation and matplotlib \cite{hunter:2007:comput-sci-eng} for data visualization. The Sage worksheet containing all computations is provided in the Supporting Information (Protocol S1) and can also be downloaded from \url{http://www.sagenb.org/home/pub/1224/} or \url{http://www.courant.nyu.edu/~ho/}.

\section*{Acknowledgments}
We thank Leslie Greengard for useful discussions and for facilitating our research. We also thank the anonymous reviewers for their very helpful comments and suggestions.

\bibliography{}

\section*{Figures}

\end{document}